\documentclass[onecolumn]{mn2e}
\usepackage{caption}
\pdfoutput=1
\usepackage{graphicx}
\usepackage{booktabs,caption,fixltx2e}
\usepackage{subfig}
\usepackage{float}
\usepackage{subfloat}
\usepackage{caption}

\begin{document}

\title[Comment]{The prediction of rotation curves in gas-dominated dwarf galaxies
with modified dynamics}

\author[R.H. Sanders] {R.H.~Sanders\\Kapteyn Astronomical Institute,
P.O.~Box 800,  9700 AV Groningen, The Netherlands}

 \date{received: ; accepted: }

\maketitle
\begin{abstract}
I consider the observed rotation curves of
12 gas-dominated low-surface-brightness galaxies -- objects in which the
mass of gas ranges between 2.2 and 27 times the mass of the
stellar disk (mean=9.4). This means that, in the usual decomposition
of rotation curves into those resulting from various mass components, the
mass-to-light ratio of the luminous stellar disk effectively vanishes as an 
additional
adjustable parameter.  It is seen that the observed rotation curves
reflect the observed structure in gas surface density
distribution often in detail.  This fact is difficult to comprehend
in the context of the dark matter paradigm where the dark
halo completely dominates the gravitational potential
in the low surface density systems;  however it is expected 
result in the context of
modified Newtonian dynamics (MOND) in which the baryonic 
matter is the only component.  With MOND the calculated 
rotation curves are effectively parameter-free predictions.
\end{abstract}
\section{introduction}

In a number of late-type dwarf galaxies, the baryonic
mass content is dominated by neutral gas
and not by luminous stars or molecular gas.  
In analysis of rotation curves in terms of various components this 
simplifies the decomposition
of the mass distribution by largely 
removing the uncertain mass-to-light ratio of the
stellar disk as an adjustable parameter.  In the
context of the standard CDM paradigm the dynamics
is dominated by the dark matter halo
which has a standard (NFW) form and the gas motion
traces, as test particles, the radial gravitational force within the halo.

Yet,  the observed rotation curves show a surprising
variety of shapes given the dominance of the 
purported halo with a standard density distribution
(Navarro, Frenk, White 1996).  This problem is
recognised  and a number of solutions 
in the context of the CDM paradigm have been proposed:
the effects of baryon-induce fluctuations due to 
gas dynamical feedback in the presence of ongoing 
star formation (Oman et al. 2015);  dynamical friction of gas clouds 
formed by gravitational instability leading to transfer of
energy to dark matter (Del Popolo et al. 2018); modification
of the presumed properties of dark matter particles so that
they are self-interacting (SIDM) and form 
isothermal halo cores (Ren et al. 2018).
Most of these mechanisms were originally designed to
solve generally perceived problems such as the 
as the observational appearance of cores in
galaxies rather than cusps as predicted in
the pure cosmological N-body calculations
(Read et al. 2016).  The problem presented by diversity of shapes 
has more recently been addressed and 
is explained by the gravitational influence of baryons on
the more responsive halo core as opposed to the
harder cusp.

But the problem is more severe than the general
issue of the diversity of rotation curves;  the form
of the observed rotation curves is
most often traced in detail by the distribution
of baryonic matter.  This is particularly true in
low-surface brightness dwarf galaxies where the
presumed dark component is overwhelmingly dominant.
Here, using a sample of 12 gas-dominated
systems, I show that this variety of shapes is
directly related to the form of the observable
gas, or effectively, the baryonic mass distribution.
The diversity of rotation curve shapes follows directly from
the diversity in the distribution of baryons, and
the relationship in form as well as amplitude is well-described 
in detail by modified Newtonian
dynamics, or MOND, proposed by Milgrom (1983) 
as an alternative to astronomical dark matter.

In this context  MOND may be viewed as an algorithm  --  a modified
Poisson relation between the observable baryonic mass
distribution and the mean gravitational
acceleration in an astronomical object.  Basically 
the idea is that the true gravitational acceleration
$g$ is related to the Newtonan acceleration $g_N$
(the derivative of a potential given by the Newtonian 
Poisson equation) by $$g\mu(g/a_0) = g_N ,\eqno(1)$$ 
where $a_0$ is a universal constant with
units of acceleration and $\mu(x)$ is an unspecified 
function that interpolates between the Newtonian 
regime ($x>>1,\, \mu(x) = 1$) and the low acceleration regime 
($x<<1,\, \mu(x) = x$).  

There are well-known observational and theoretical problems with 
non-relativistic MOND on scales
larger than galaxy groups (but see Milgrom 2018):  
not accounting for the full
discrepancy in rich clusters of galaxies;  absence of
a consistent cosmology or cosmogony;  no definite
prediction of the form of anisotropies in the 
cosmic background radiation -- all problems that
are addressed by the paradigm of cold dark matter.
MOND still requires a consistent relativistic extension, but
the fact that  $a_0 \approx cH_0/6$ is suggestive of
a cosmological connection.  While these are issues
that must be addressed by a full theory, they do not
subtract from the success of MOND in the treatment of
galaxy rotation curves (see Famaey and McGaugh 2012
for an up-to-date review of the theoretical status of MOND).

An advantage of considering
this sample of dwarf galaxies is that the objects are
mostly in the low acceleration regime and the exact form
of the interpolating function does not play a significant role, 
i.e. $g \approx \sqrt{g_N a_0}$.
Moreover, the distribution of the baryonic surface
density, the gas, is directly observed.  Thus, the fact that
the mass-to-light ratio of a stellar component  
vanishes as an adjustable parameter means that
the MOND rotation curves are essentially
parameter-free predictions:
{\it{MOND predicts the form and amplitude of the rotation 
curves from the observed distribution of baryons 
with only one additional universal parameter 
having units of acceleration.}}    And it 
works well in most cases.
 
The prediction of rotation curves from the observed distribution
of baryonic matter is extremely challenging to the dark matter
paradigm because it is not evidently a property
permitted by a non-interacting, non-dissipative medium
(also SIDM) .  Until any dark
matter model can achieve comparable predictive success
with one additional fixed parameter, the CDM model 
cannot be considered on a par with MOND on the scale of
galaxies.

\section{The Sample}

The objects considered are listed in Table 1.
Seven of these are from the HI observations of 
dwarf galaxies, the "little things"  (LT) survey (Hunter et al. 2003).  
These are galaxies with
maximum observed rotation velocities in the
range of 30 to 60 km/s and gas (neutral hydrogen 
and helium)  to stellar mass ratios ranging from  
five to 30; in other words, they are extremely gas dominated
systems.  The remaining five are from the literature
including three from the sample of dwarfs described
by Swaters et al. (2010)  These also are
highly gas dominated objects ranging up to ratios
in excess of 20.

For the dwarf galaxies there are well-known interpretive difficulties 
presented by the use of two-dimensional velocity fields to 
the determination of the run of circular velocity. 
These galaxies are, after all, irregular
which means that they are generally asymmetric
in both their light and gas distributions
as well as their kinematics.  For the
LT galaxies these problems are particularly severe. 
The asymmetry results in a basic  uncertainty
in the projection of the measured density or velocity 
fields as embodied primarily by the inclination 
parameter (Oh et al. 2015).

By what factors should the velocity fields be 
de-projected?  Typically, in such analyses, the average inclination 
is restricted to be greater than 50 but less than
80 degrees to reduce ambiguities resulting from
de-projection.  But only 15 of the objects in the  LT
sample of Oh et al. meet this criterion in terms of their 
adopted global inclinations.  Moreover,
there is also a large scatter of fitted inclinations 
in the tilted ring analysis of single objects (see DDO 70
for example, given in Oh, et al,).  The irregular light and gas 
distributions, and
in several cases substantial differences between
the photometric and kinematic inclinations,
suggest the presence of basic structural asymmetries (bars for example)
that can lead to large scale and un-modelled 
non-circular motions.  
We should bear in mind that all of these effects can
call into question the role of the published rotation
curves as an accurate tracer of the underlying gravitational 
force.

To minimise these difficulties the seven LT galaxies included 
here all satisfy the following criteria:  
1) all global inclinations lie between 40 and 80 degrees;  2) the 
difference between
the kinematic and visual inclinations is less than 10 degrees
(this reduces the possibly significant contribution of bars to 
the velocity fields);
objects with strongly asymmetric velocity fields 
(see Iorio et al. 2017) are not included 
(this is also indicated by a highly variable and fluctuating 
inclination parameter 
in the tilted ring algorithm).  These seven are combined with
five additional gas-rich galaxies from the literature to form
the total sample.

In what follows I fix all distances, inclinations and
stellar masses at
their nominal values in the given references.  For example,
in all of the LT galaxies these parameters are taken as given
in Oh et al.   For the UGC dwarfs the values are taken 
from Swaters et al.  This  means that there is no
tweaking of parameters to achieve a rotation curve that
more closely agrees with the observations.  Thus the procedure
is more in
the spirit of prediction and not fitting, but one should 
keep in mind that the expectations should be somewhat
lower than in the case where parameters may slide.

The baryonic mass-rotation velocity relation (Tully-Fisher)
described by these gas-rich dwarfs is shown in Fig.\ 1. 
Here solid line is not a fit but is
the relation predicted by MOND, $v^4 = Ga_0M_b$,
where $a_0  = 1.2\times 10^{-10}$ m/s$^2$ as usual.
This immediately demonstrates that the amplitude of 
the rotation curves 
is directly related to the mass of baryonic matter
in the sample gas rich galaxies in the same manner 
as in objects with higher rotation velocities with a 
mass distribution dominated by the stellar disk (McGaugh 2005),
and this indicates that the distances are not too far in error.
(not by 25\% for example).

\begin{figure*}
\begin{centering}
\includegraphics[height=50mm]{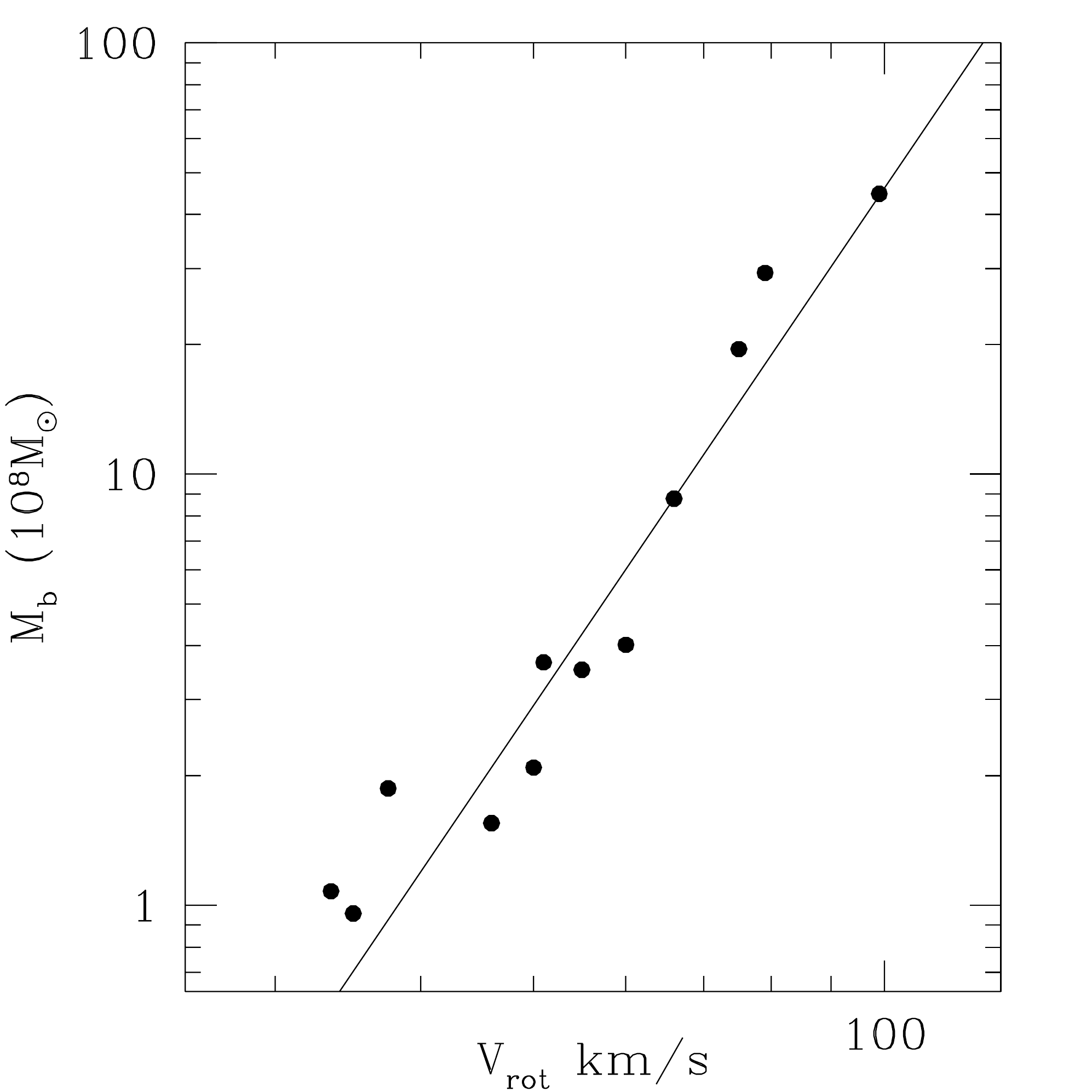}
\caption{The baryonic Tully-Fisher relation defined by the
sample of gas rich dwarfs.  The contribution of the stellar
mass is included as specified in Table 1.  The line is 
the MOND prediction ($V^4=Ga_0M$ with
$a_0= 1.2\times 10^{-10}$ m/s$^2$).}
\end{centering}
\end{figure*}

\section{Predicted rotation curves}
\begin{table*}
\center
\caption{Gas Dominated Galaxies}
\label{symbols}
\begin{tabular}{@{}lcccccccc}
\hline
Galaxy & $D$ & $V_{rot}$ & Incl. kin. & Incl. ph.
& $M_{gas}$
& $M_{stars}$  &  $M_{gas}/M_{stars} $\\ &  $Mpc$
      &  km/s &  Degrees & Degrees & $10^8$ & $M_\odot$ & $10^8$ $M_\odot$
      & refs. D \& Incl. \\
\hline
 DDO 52 & 10.3 & 60 & 43 & 51&  3.3 &  0.72 & 4.6 & 1 \\
 DDO 87 & 7.7 & 55 & 56 & 59 & 2.9 & 0.62 & 4.7 & 1 \\
 DDO 126 & 4.9 & 38 & 65 & 68 & 1.6 & 0.23 & 7.2  & 1 \\
 DDO 133 & 3.5 & 46 &  43 & 49 & 1.3 & .26 & 4.9  & 1 \\
 DDO 154 & 3.7 & 51 & 68  & 65 & 3.5 & 0.13 & 27. & 1 \\
 IC 2574  & 3.2 & 66 & 77 & 83 & 8.1  &  0.67  & 16. & 2 \\
 NGC 3741 & 3.2 & 50 & 64 & 57 & 2.0. & 0.087 & 23. & 3 \\
 Har 29  & 5.9 & 34 & 61 & 59 & 0.94  & 0.14 & 6.5  & 1 \\
 UGC 4499 & 13.9 & 75 & 50 & 47  & 15.0 & 4.5 & 3.1 & 4 \\
 UGC 5005 & 53 & 99 & 40 &41 & 44. & 4.7 & 8.5 & 4  \\
 UGC 5750 & 59 & 79 & 61 & 70 & 20. & 9.3 & 2.2& 4 \\
 WLM &     1.0 & 35 &  74 & 70 & 0.80 & 0.16 & 4.9 & 1 \\
\hline
\end{tabular}
\item{1) Oh, et al. 2015; 2) Sanders 1996; 3) Gentile et al. 2007; 4)  Swaters, et al.  718.}
\medskip
\end{table*}

In calculating
predicted rotation curves I take the Newtonian rotation curves of the
baryonic components, assumed to be in a thin disk, as given in the
indicated references (Table 1, column 9).  I convert to the MOND rotation
by applying eq.\ 1 where the the interpolating function is assumed
to be of the ``standard" form ($\mu(x) = x/\sqrt{1+x^2}$). As noted above
for the low surface density objects (low acceleration) the
calculated rotation curves are relatively insensitive to the exact form.
(in all cases, the calculated rotation velocities are within a few
percent of that given by the asymptotic form of $\mu(x)\approx x$).

For the sample galaxies. the resulting rotation curves predicted by 
Newton (dashed curves) and by MOND (solid curves)
are compared with the observations in figures 2 (lower panels), 
along with the surface density
distributions of the baryonic components.  It is striking that in several
cases (DDO 87, WLM, IC 2574, NGC 3741) the observed structure corresponds
quite precisely to that predicted by the baryonic mass distribution
in the context of MOND.

\begin{figure}
  \begin{minipage}[b]{0.6\textwidth}
    \includegraphics[height=11cm]{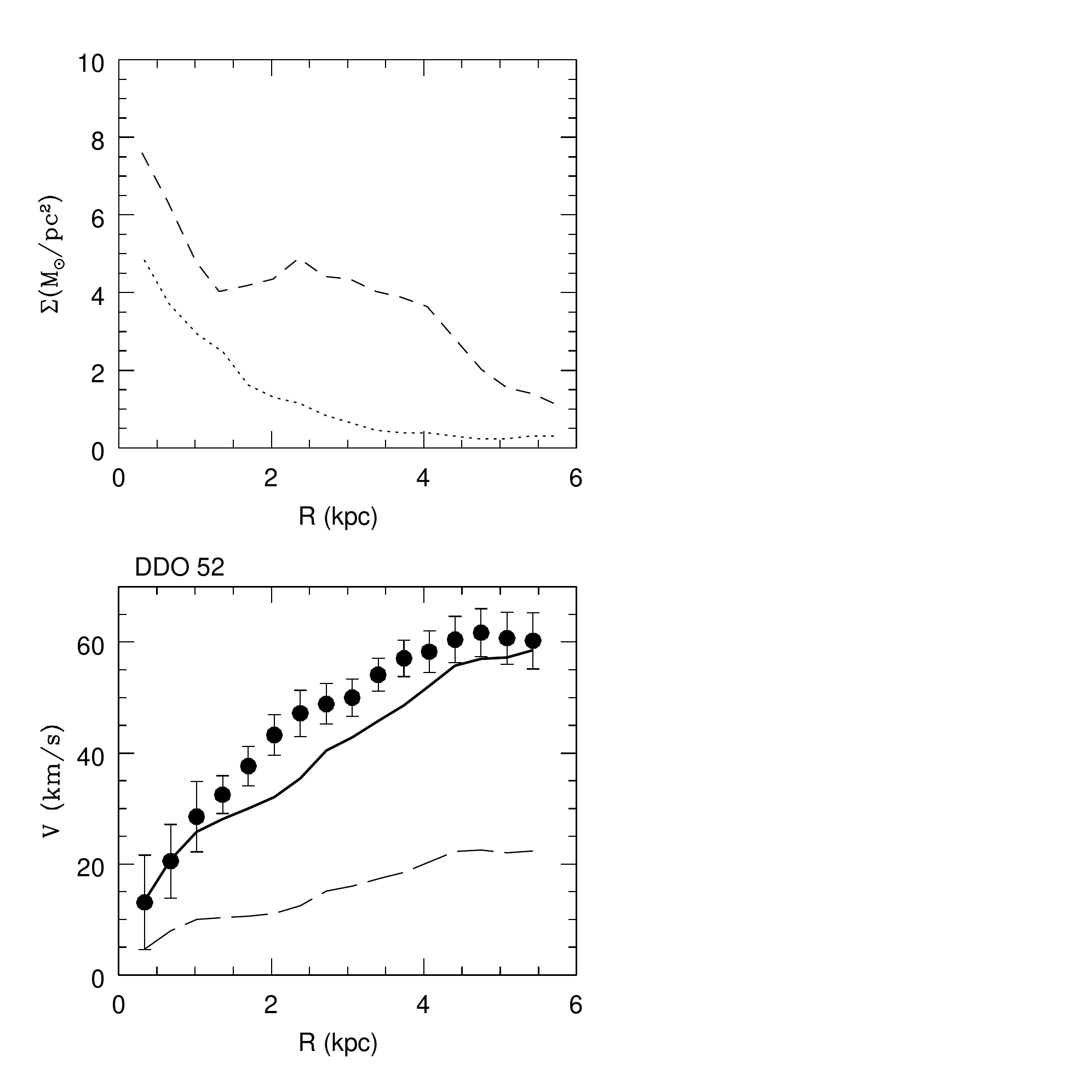}
    \captionsetup{labelformat=empty}
    \end{minipage}
  \begin{minipage}[b]{0.6\textwidth}
    \includegraphics[height=11cm]{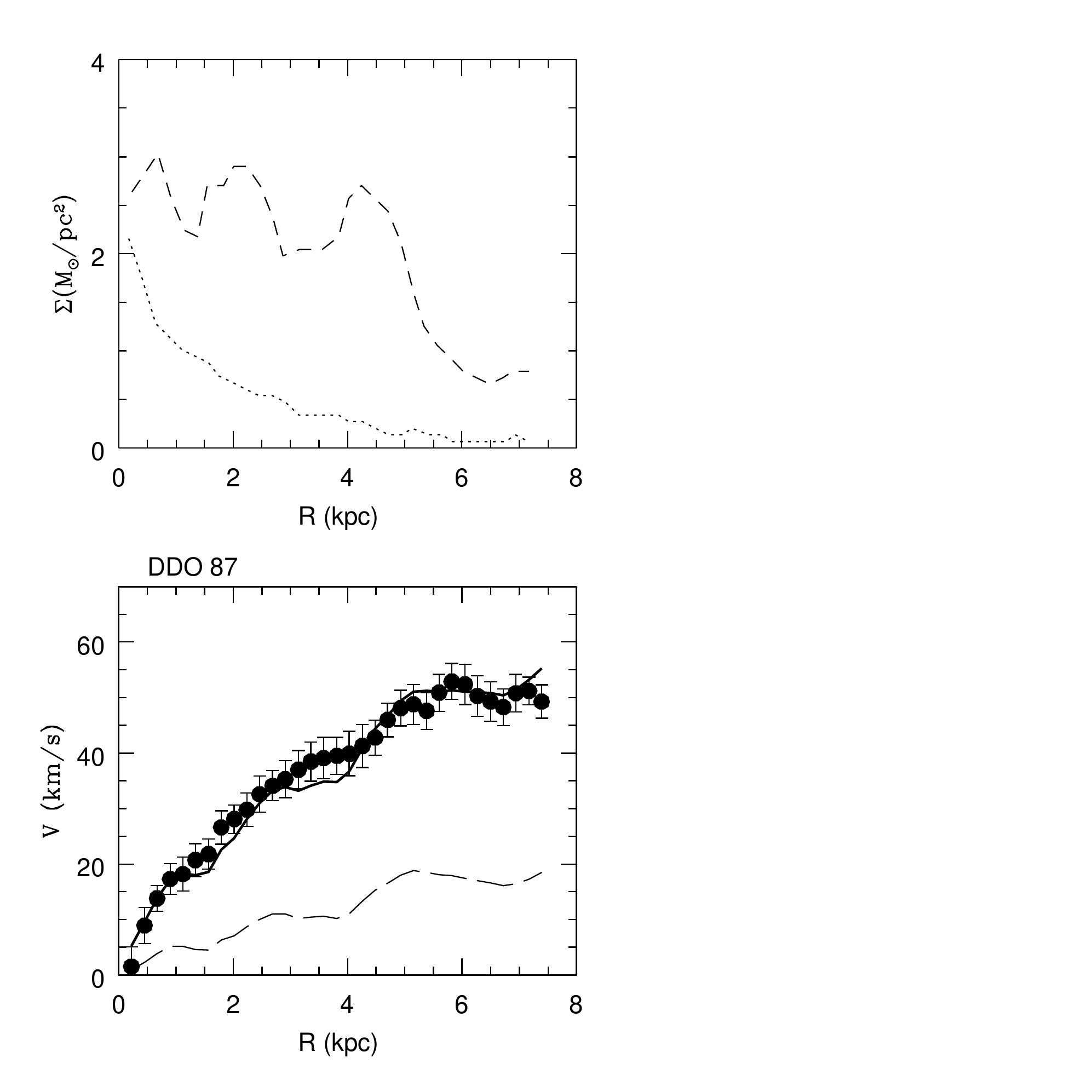}
        \captionsetup{labelformat=empty}
  \end{minipage}

  \begin{minipage}[b]{0.6\textwidth}
    \includegraphics[height=11cm]{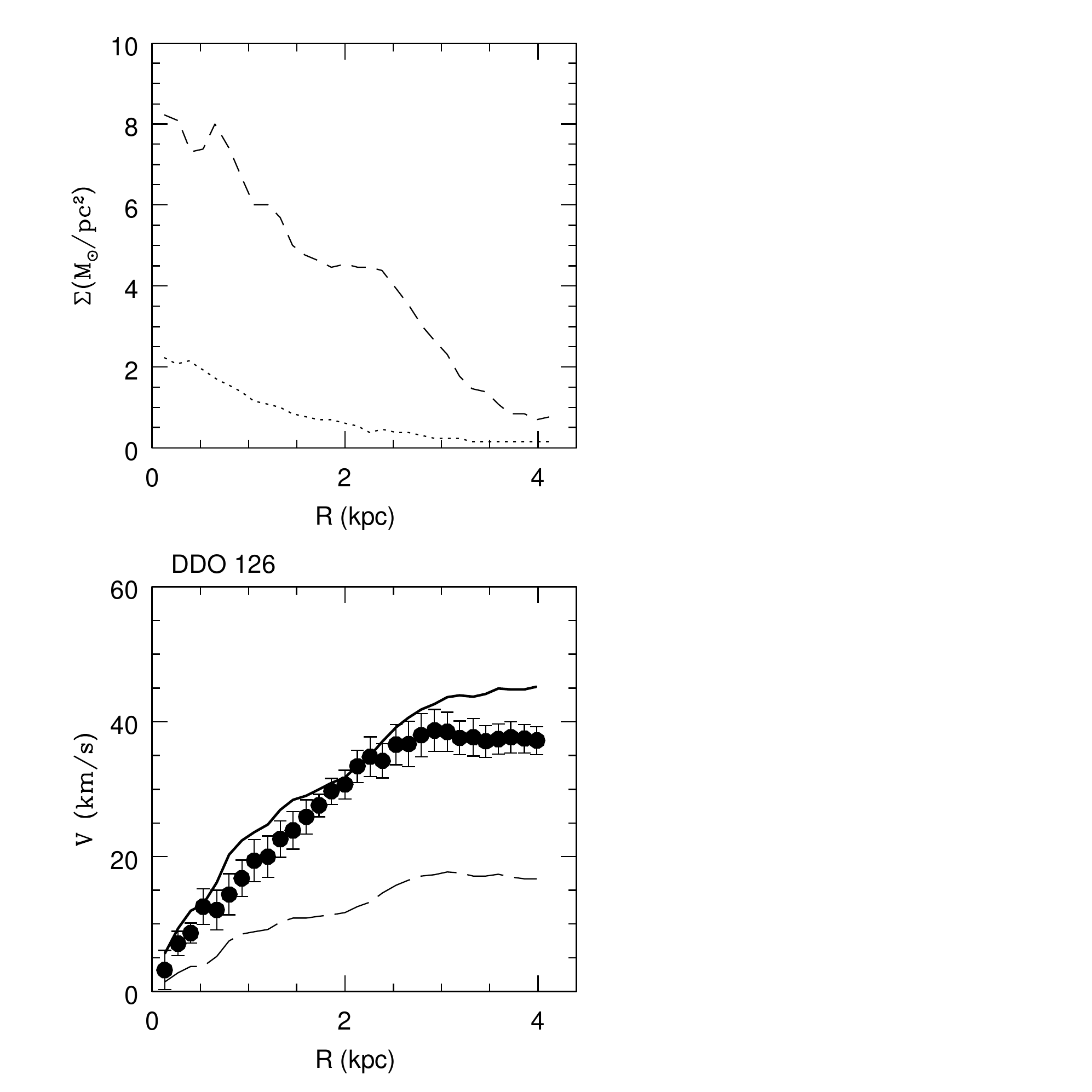}
        \captionsetup{labelformat=empty}
  \end{minipage}
  \begin{minipage}[b]{0.6\textwidth}
    \includegraphics[height=11cm]{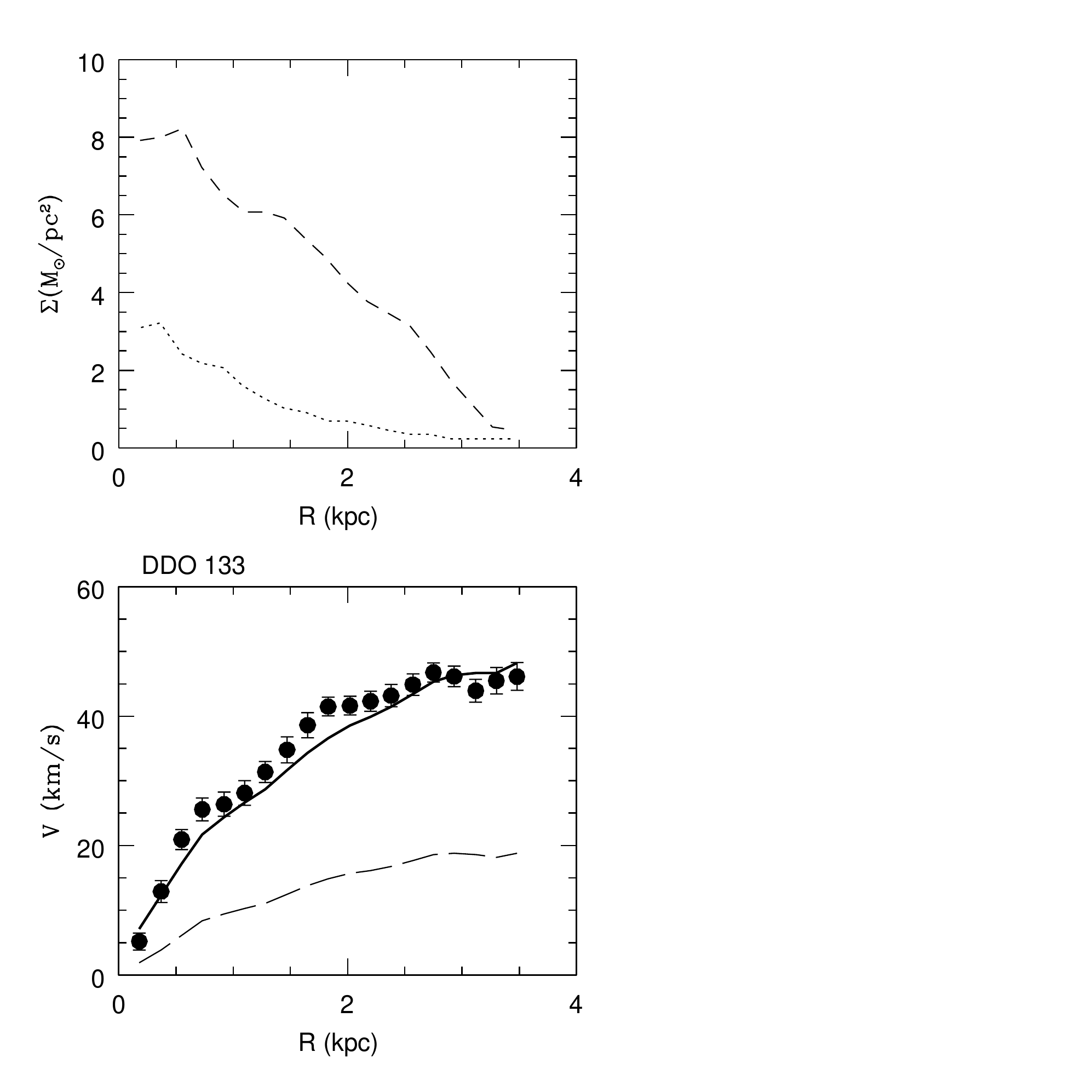}
        \captionsetup{labelformat=empty}
  \end{minipage}
  \end{figure}
\vfill

\begin{figure}
  \begin{minipage}[b]{0.6\textwidth}
    \includegraphics[height=11cm]{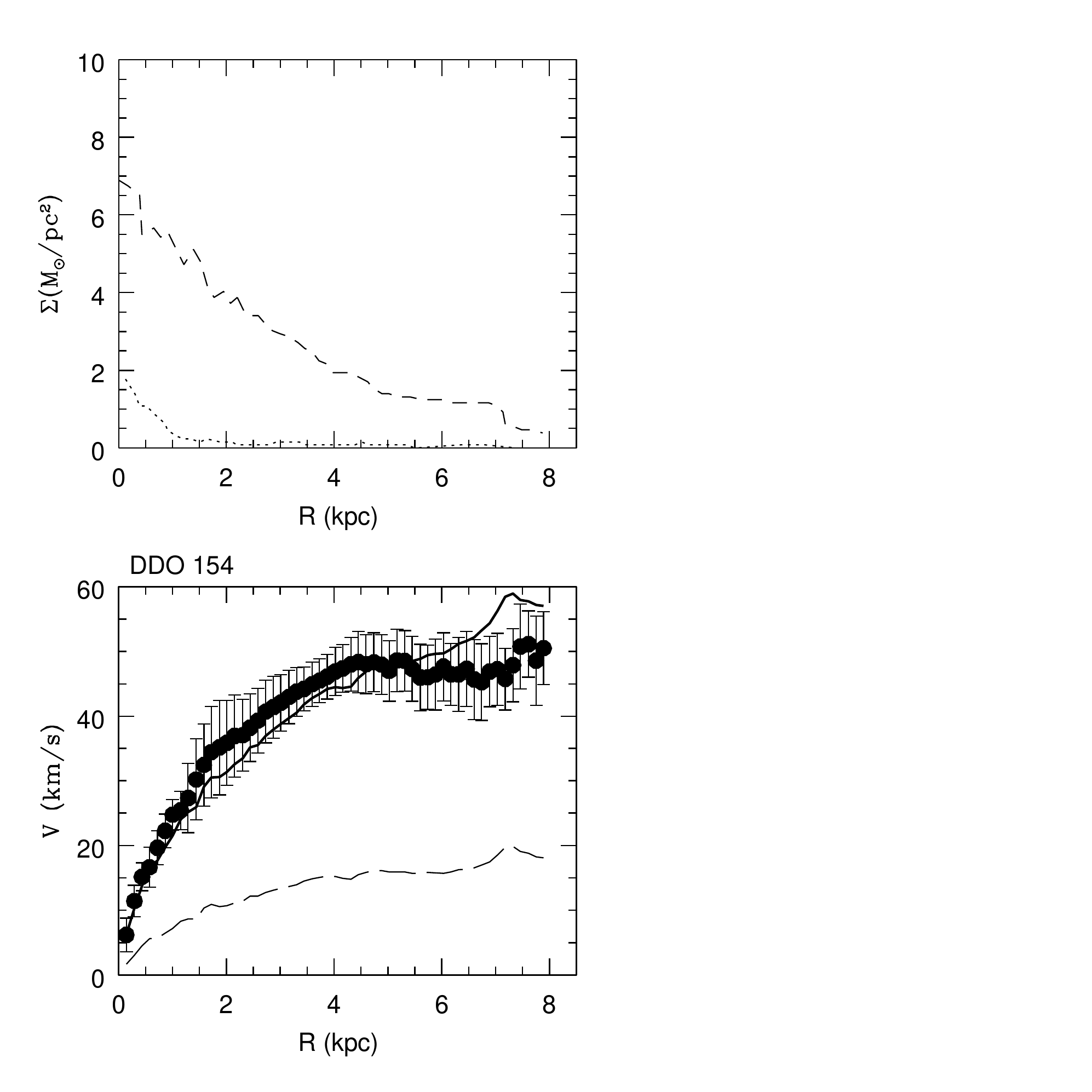}
        \captionsetup{labelformat=empty}
  \end{minipage}
  \begin{minipage}[b]{0.6\textwidth}
    \includegraphics[height=11cm]{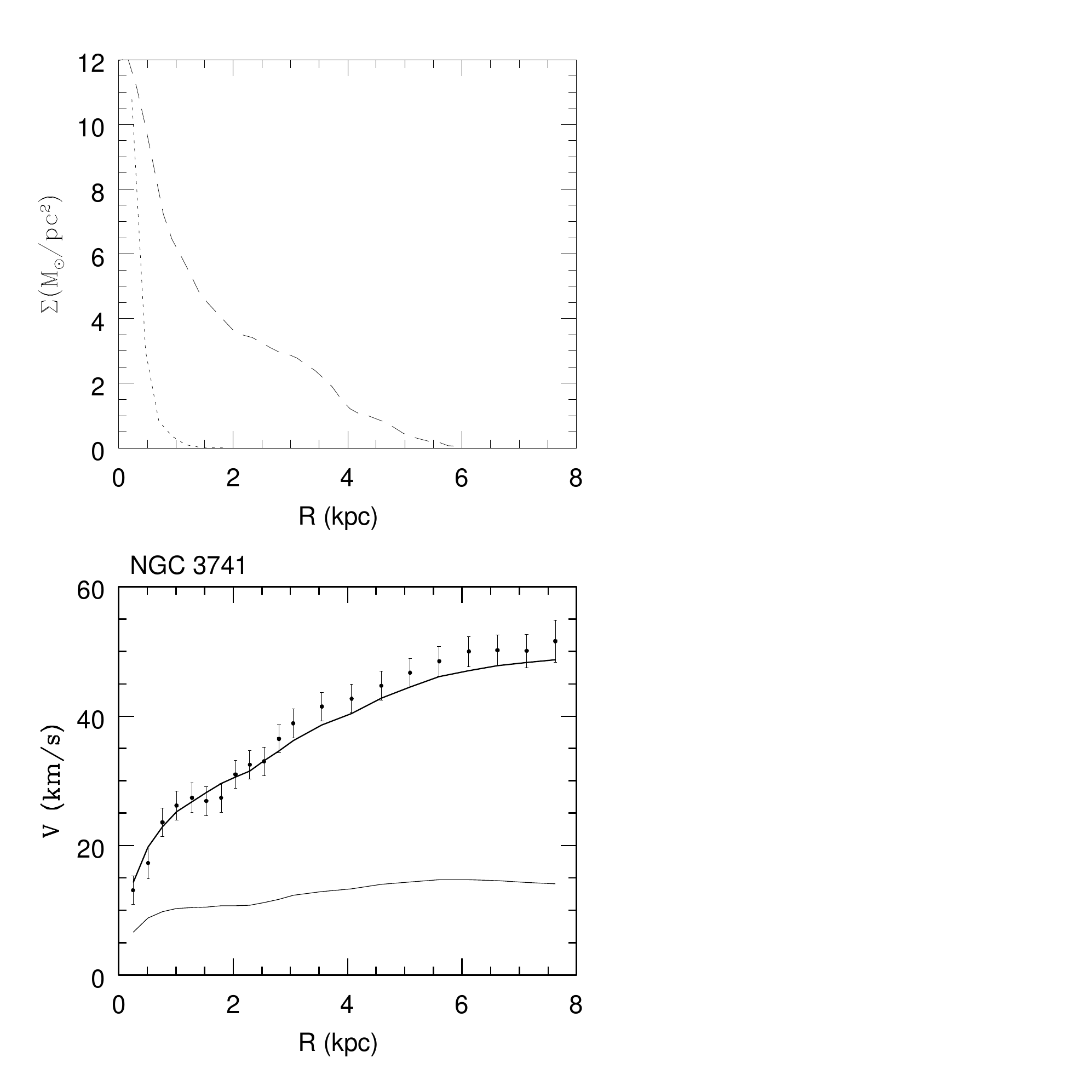}
        \captionsetup{labelformat=empty}
  \end{minipage}

  \begin{minipage}[b]{0.6\textwidth}
    \includegraphics[height=11cm]{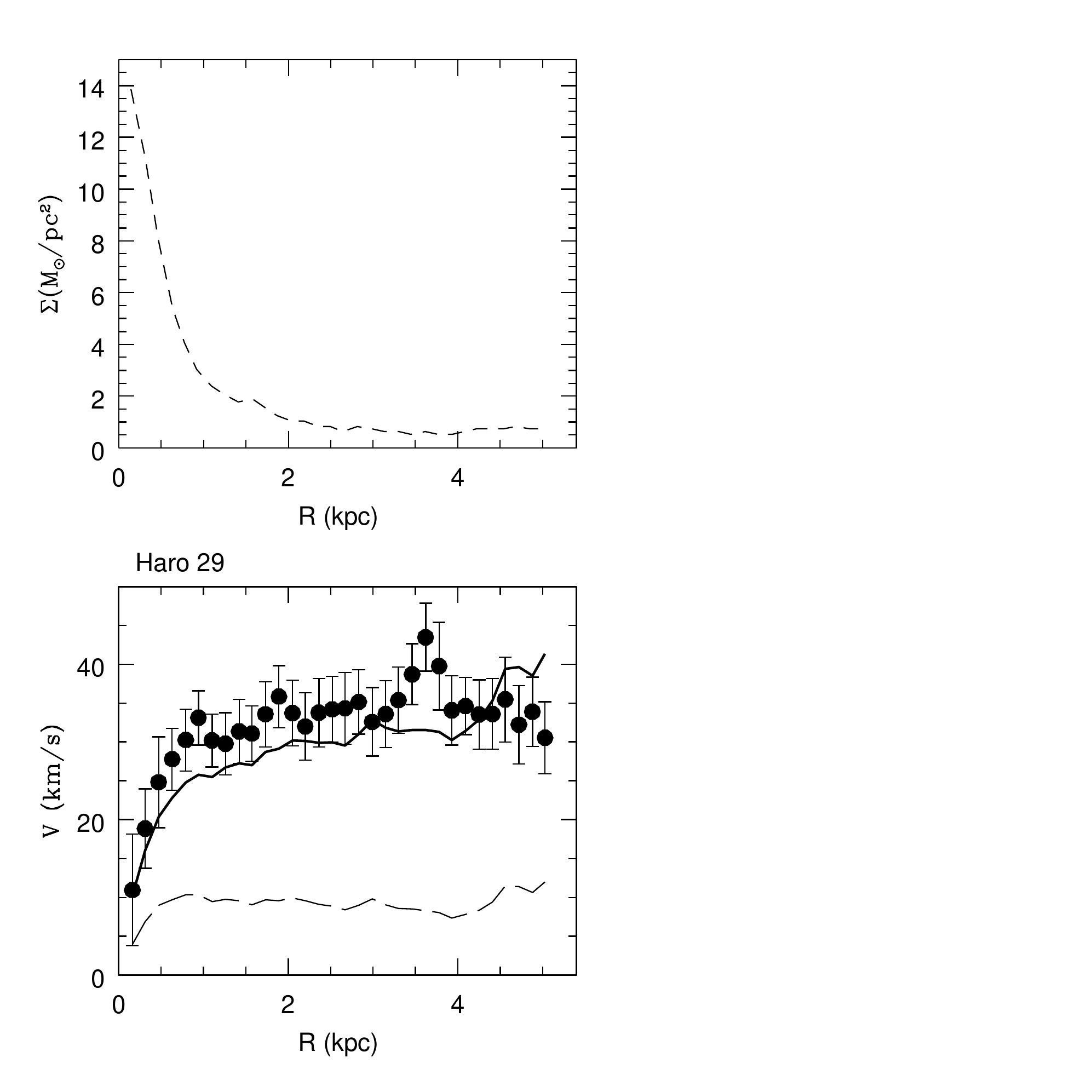}
        \captionsetup{labelformat=empty}
  \end{minipage}
  \begin{minipage}[b]{0.6\textwidth}
    \includegraphics[height=11cm]{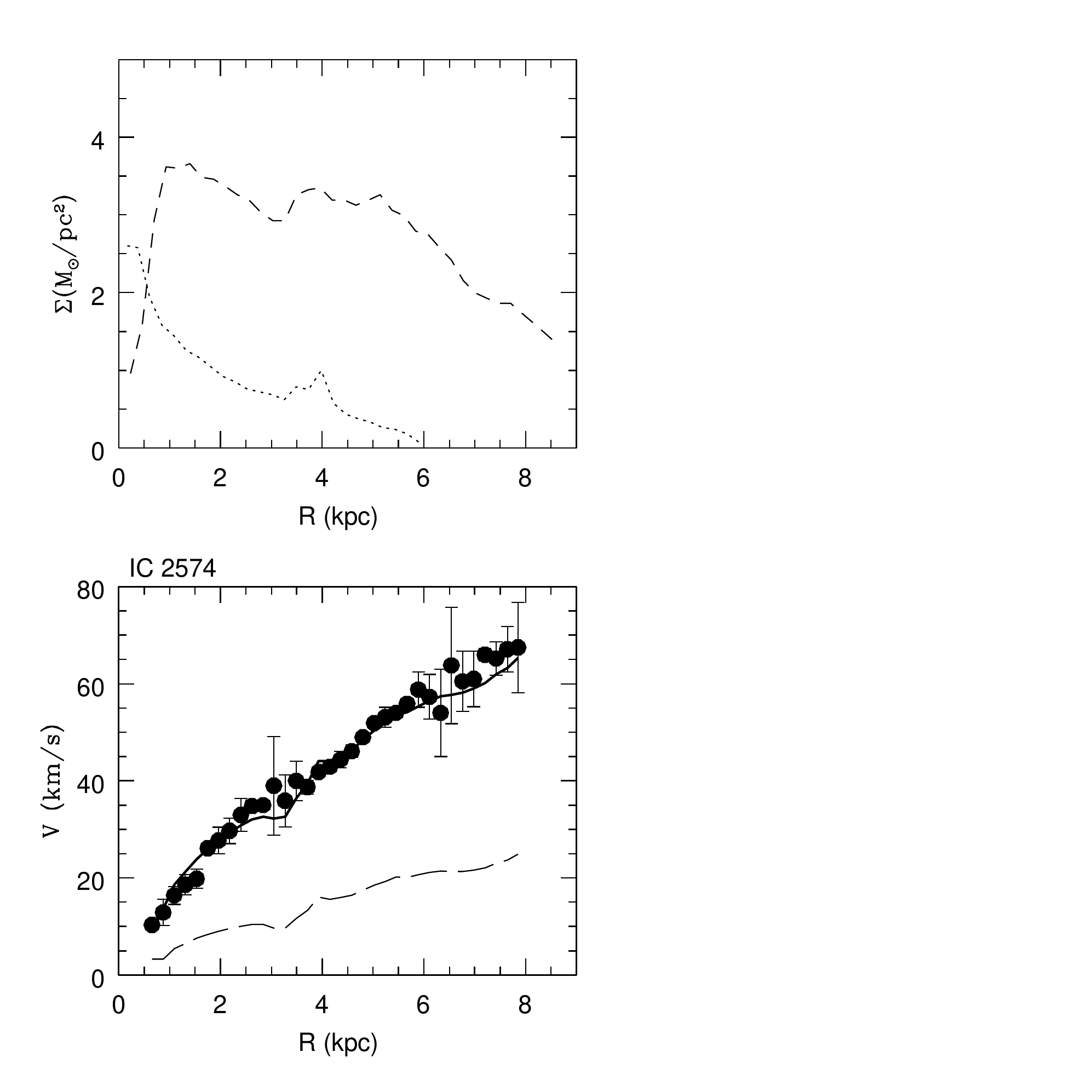}
        \captionsetup{labelformat=empty}
  \end{minipage}
  \end{figure}
  \vfill

\begin{figure}
  \begin{minipage}[b]{0.6\textwidth}
      \captionsetup{labelformat=empty}
     \includegraphics[height=11cm]{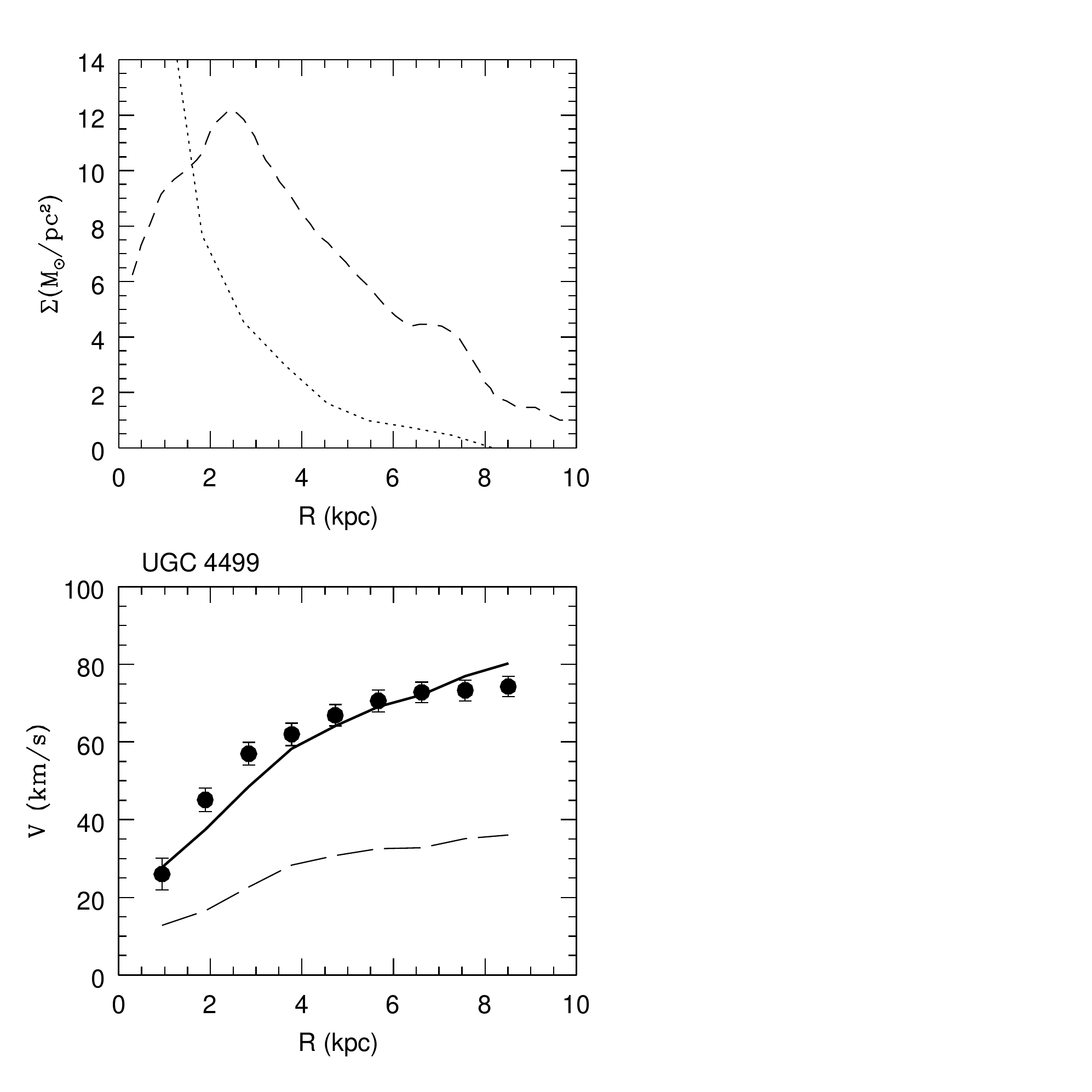}
         \captionsetup{labelformat=empty}
  \end{minipage}
  \begin{minipage}[b]{0.6\textwidth}
    \includegraphics[height=11cm]{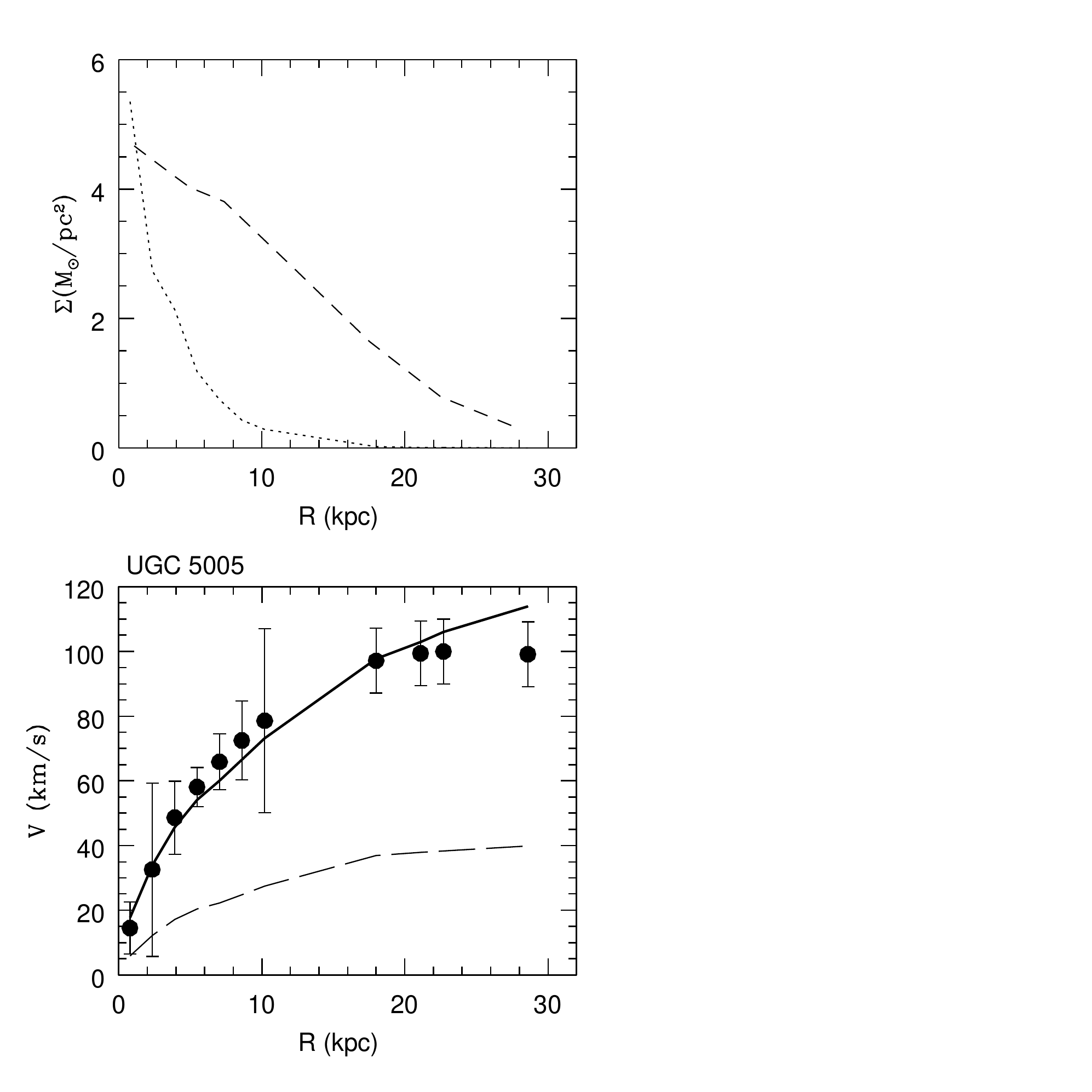}
        \captionsetup{labelformat=empty}
  \end{minipage}

  \begin{minipage}[b]{0.6\textwidth}
    \includegraphics[height=11cm]{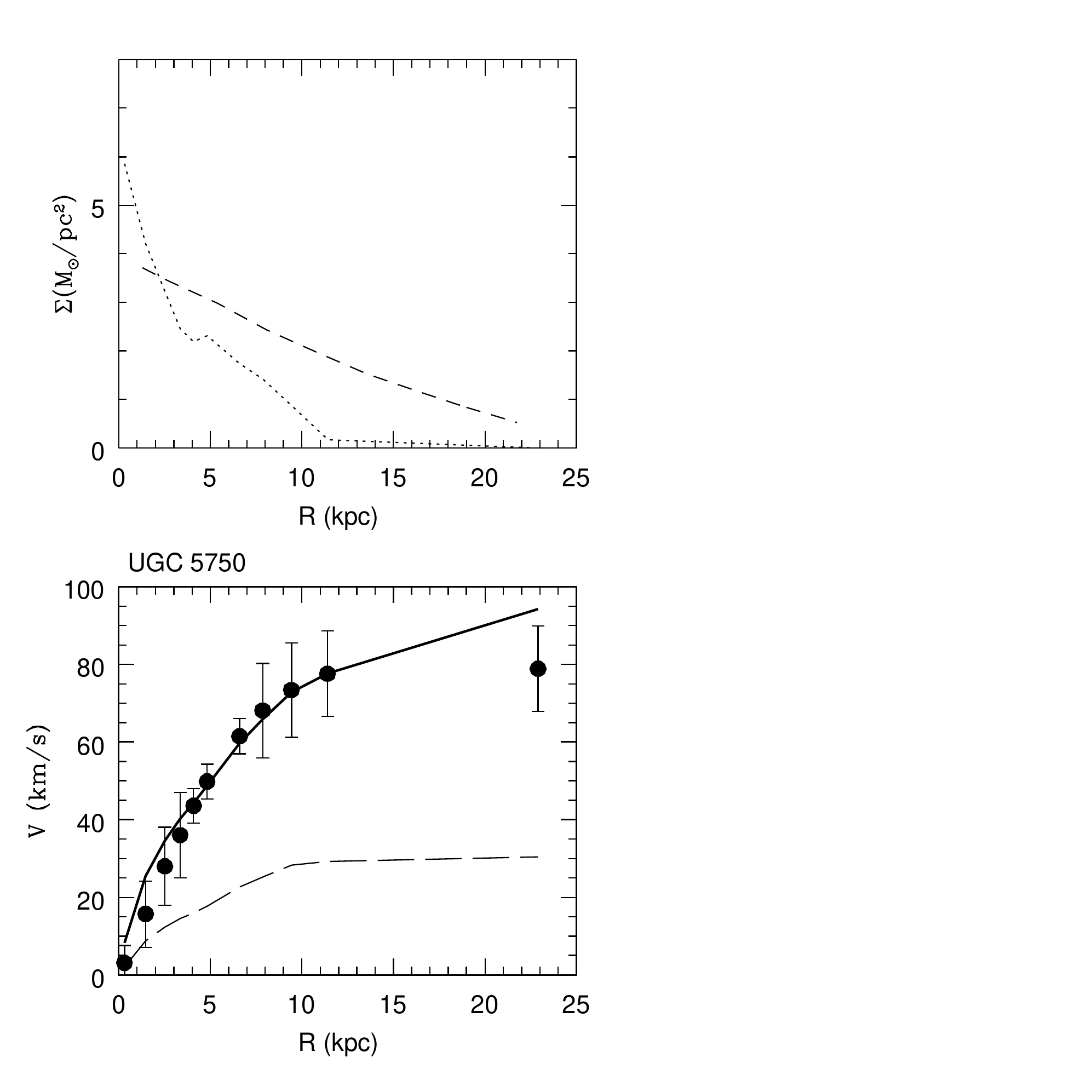}
        \captionsetup{labelformat=empty}
  \end{minipage}
  \begin{minipage}[b]{0.6\textwidth}
    \includegraphics[height=11cm]{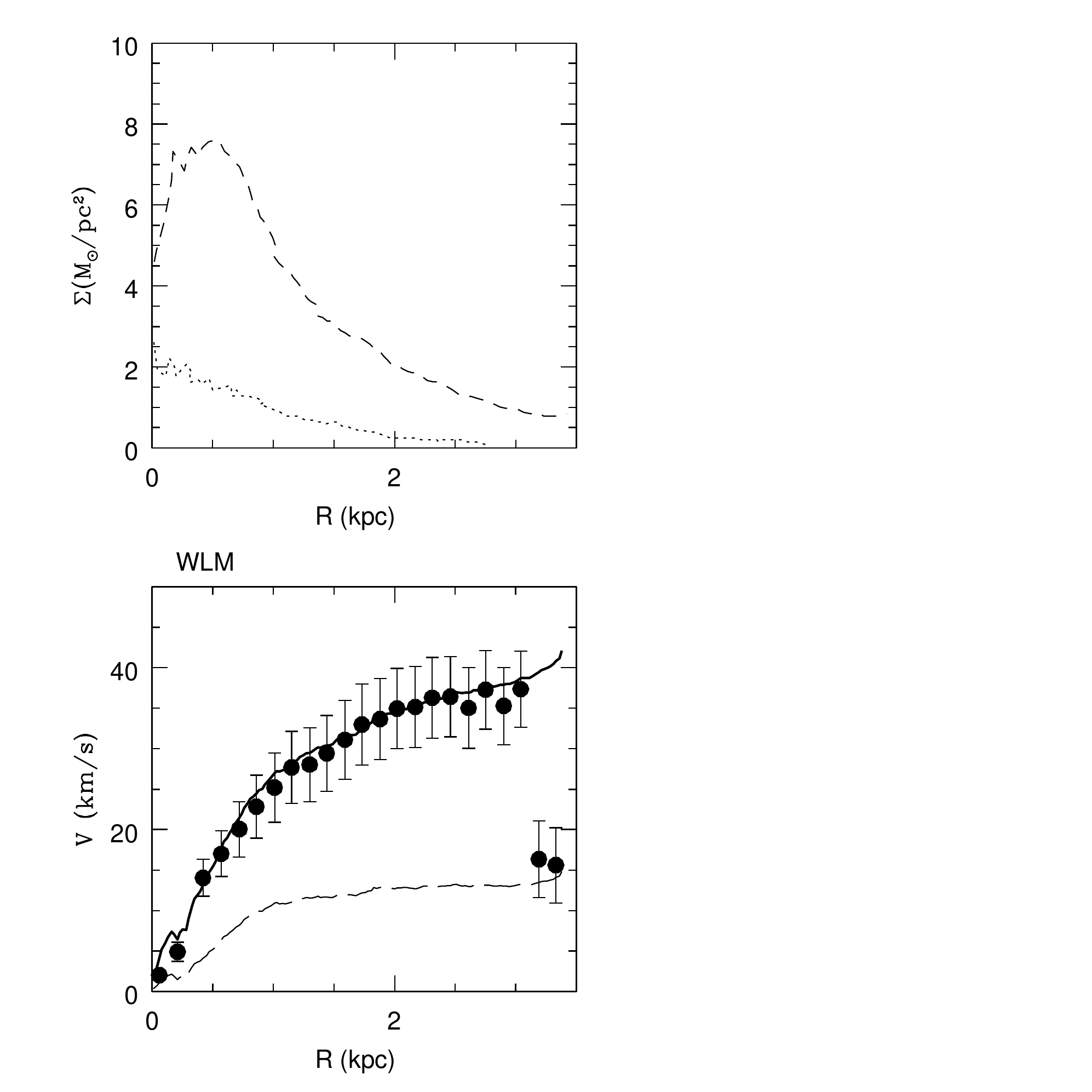}
        \captionsetup{labelformat=empty}
  \end{minipage}
\caption { Each galaxy is represented in two panels.  The upper panel is the  
    surface densities of gas
    (dashed) and starrs (dotted) as a function of radius.  The lower panel show the 
    observed
    rotation curve (points), the Newtonian rotation curve of
    baryonic components (long dashed curve , and MOND rotation curve.
    calculated from the Newtonian curve via eq.\ 1 (long dashed curve).}

\end{figure}

\vfill

\begin{figure*}
\begin{centering}
\includegraphics[height=60mm]{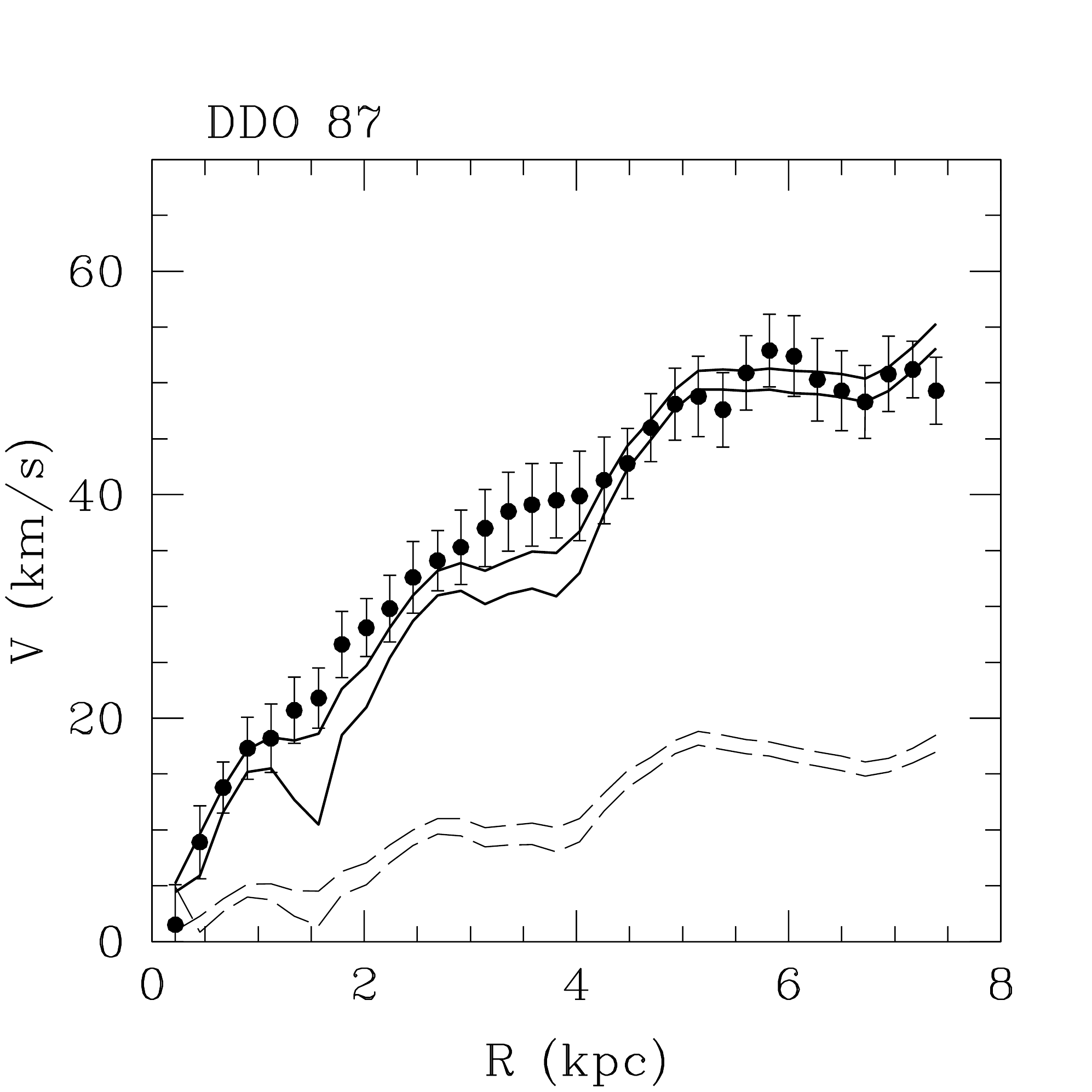}
\caption{Predicted rotation curves of DDO 87, with and without stars.  The dashed 
lines are the Newtonian rotation curves and the solid curves
are the MOND rotation curves.  For both Newton and MOND the lower curves
are with only the gaseous component and no stellar component (17.5 \% of total baryonic mass).
The upper curves are calcuated with the total baryonic mass.}
\end{centering}
\end{figure*}

The claim here is that these rotation curves are essentially predictions
with no free parameters because of the dominance of the gaseous
component.  But of course there is some contribution from the
the luminous disk so one might question the extent to which 
M/L of the disk does vanish as an adjustable parameter.
The stellar masses of the sample galaxies (column 6 in table 1)
are taken from fitting the colors of the galaxies to population
synthesis models in the cases of the seven LT galaxies (Oh et al. 2015).  
For the other five cases they are taken from kinematic considerations
(e.g., not exceeding the observed rotation velocity).
In half the cases, the implied stellar masses exceed 15\% of
the total baryonic mass (those objects in which the ratio of gas mass to
stellar mass is less than six, column 8).  This can certainly affect
the shape of the galaxy rotation curves, especially in the inner
regions where the stellar disk often contributes more significantly
to the baryonic surface density distribution.  

\begin{figure*}
\begin{centering}
\includegraphics[height=60mm]{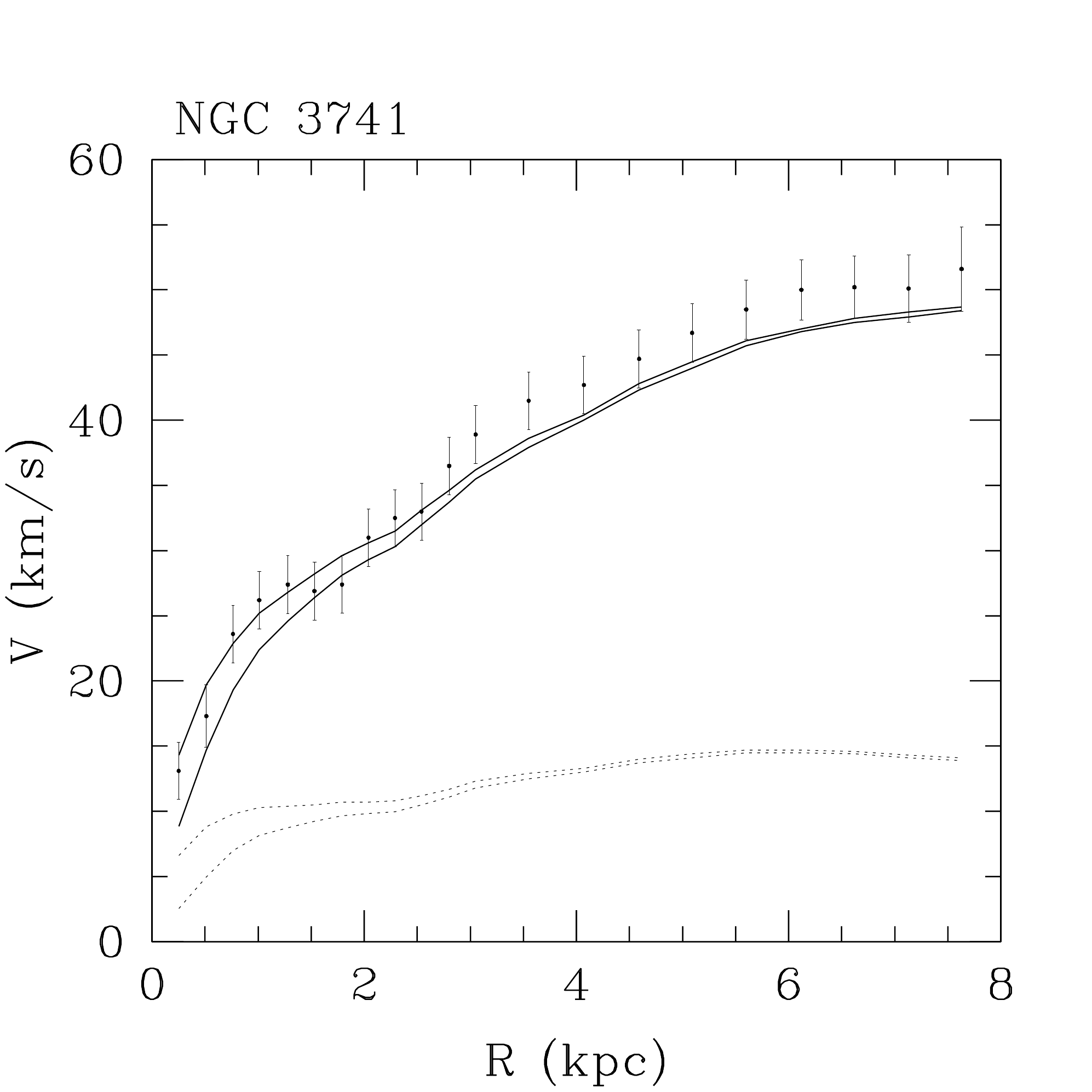}
\caption{Predicted rotation curves of NGC 3741 with and without 
stars.  As above, the dashed 
lines are the Newtonian curves and the solid curves
are the MOND rotation curves.  In both cases the lower curves
include only gas with no stellar component (4 \% of total baryonic mass) and
the upper curves are calculated with the total baryonic mass.}
\end{centering}
\end{figure*}
\vfill

In Fig. 3  we see the calculated rotation
curves for one case where the fraction of stellar mass is 
fairly typical, DDO 87.  The lower dashed curve is
the Newtonian rotation curve  of baryons excluding the estimated
contribution of the visible disk and the upper dashed curve is
the same but including the stellar disk with a contribution
of 17.5\% of the total baryonic mass as given in
table 1;  the upper and lower
solid curves are the same but for the MOND rotation curves.
While it is obvious that the the contribution of the stellar
disk provides better agreement with the observations, 
it is also clear that, with no stellar contribution at all, the 
rotation curve is qualitatively similar:  The overall dependence
of the rotation curve shape is quite independent of the 
mass-to-light ratio of the stellar disk and the final asymptotic
rotation velocity given by MOND is essentially identical with
and without stars. 

The case of an extremely gas rich galaxy, NGC 3741, is shown
in Fig, 4 where again we see the calculated rotation curves
with and without the contribution of the stellar disk compared
to the observations (Gentile et al. 2007).  For this object the gas disc 
extends 40 exponential scale lengths beyond
the luminous disk 
and is 23 time more massive;  i.e., the 
stellar disk is roughly 4\% of the total baryonic mass and, based
upon $M/L_K = 0.3$, is consistent with stellar population synthesis.
The calculated rotation curves, with and without stars, agree in the outer
regions but deviate somewhat in the inner regions where the
stellar disk contributes.  The observed rotation curve
suggests the presence of two components, and in the MOND
prediction these are the two baryonic components:  the central luminous
disk and extended neutral gas disk. But the overall agreement 
of the calculated rotation curves with and without stars demonstrates
the independence of the calculated rotation curve 
from the M/L of the stellar disk.  

\begin{figure*}
\begin{centering}
\includegraphics[height=60mm]{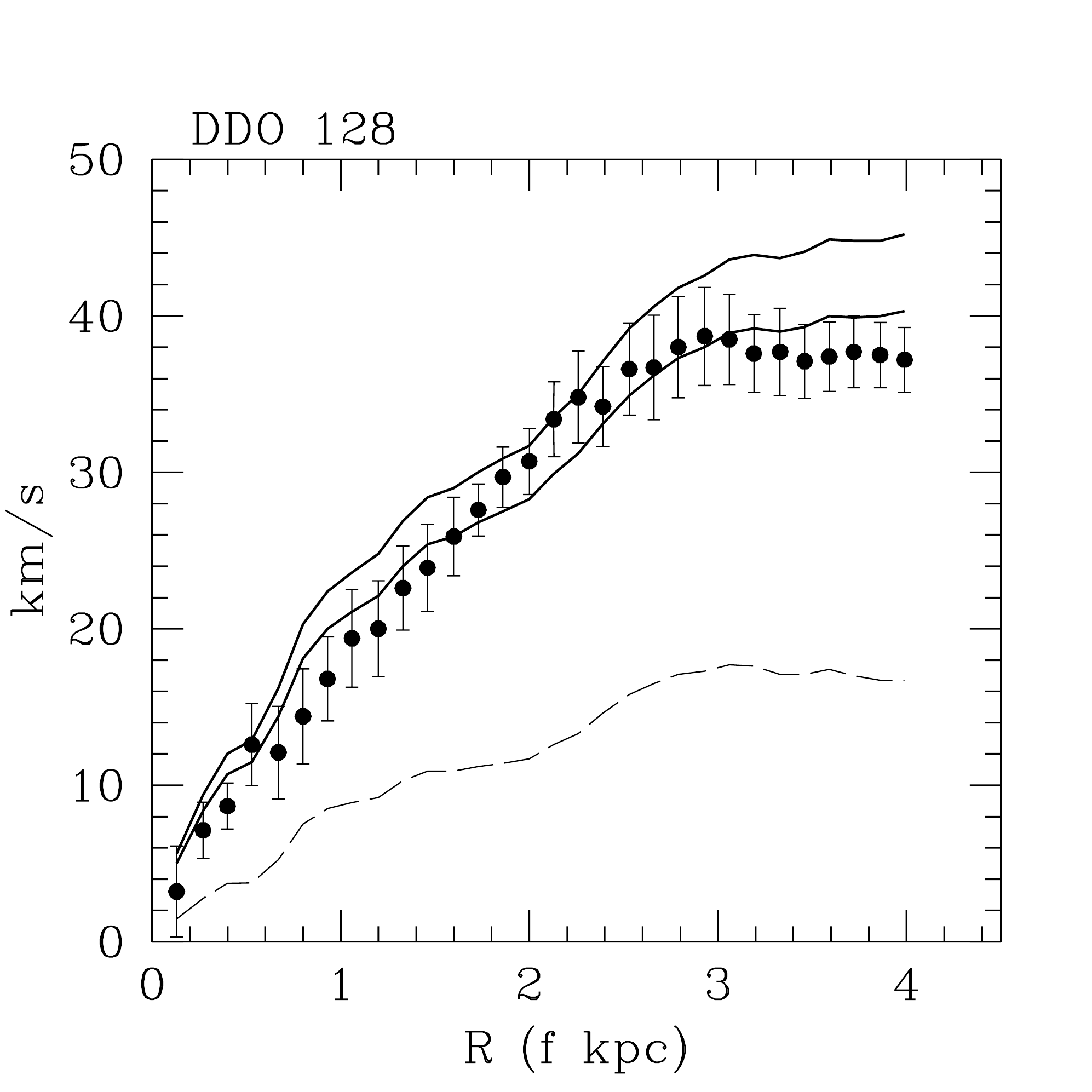}
\caption{The effect of distance estimates on the predicted rotation curve
of DDO 126.  Solid curves are the MOND curves fpr two different
distance estimates:  the upper curve for the distance adopted by Oh et al.
(4.9 Mpc) from Karachentsev et al. (2003).  The lower solid curve is taking the lowest estimate
of the distance
referenced by Karachentsev et al. (3.9 Mpc).  For this lower
value the units of radius (x axis) are scaled by the ratio of the distances
(f=0.8).  The dashed curve is the Newtonian curve of the baryonic components.}
\end{centering}
\end{figure*}

An additional factor that can affect the prediction
of rotation curves is the uncertainty of the distance 
estimate --  particularly true in an acceleration-based
modification such as MOND.  For most of these objects,
in particular those closer than 10 Mpc, the distance is
estimated via the ``tip of the red giant branch" method (TRGB)
which is generally thought to be accurate  to within 
5\%.  Therefore I take these distances as given
in the literature, primarily by Hunter et al. (2012) for the
LT galaxies.  But one should bare in mind that there can
be systematic differences in the distance scaling 
between these nearby
objects and more distant objects with Hubble law 
determinations.  Moreover, there are indications that
the reported accuracy may be optimistic.  For example,
in the case of DDO 126 the estimated distances range
from 3.9  to 5.1 Mpc (Karachentsev et al. 2003), 
all of these relying upon the
same method (TRGB)  So one should be cautions
in taking the reported errors too seriously, particularly 
for these nearby objects. 

For  DDO 126 the shape is reasonably well matched but
the predicted amplitude is too high.  Given the range of estimates 
for the distance  we see that 
this could be due to an overestimate of
the distance to the galaxy.  Figure 5 shows the predicted rotation curves
for the preferred distance of Oh et al.,  4.9 Mpc, (upper solid curve) as well as the
lowest value cited by Karachentsev et al., 3.9 Mpc (the lower solid curve).   
In the case of the shorter estimate the agreement of the MOND prediction with
the observed curve is more precise.
For these galaxies in general
predicted rotation curve agrees with that observed to within the likely uncertainty
introduced by the distance and inclination.
This illustrates that the given error bars do not capture the systematic 
effects introduced by the uncertainty in the fundamental parameters of distance
and projection.

\section{Conclusions}

MOND predicts an observed phenomenology in
galaxies that has only recently been considered
by dark matter theorists.  For example, there is the
``radial acceleration relation" (RAR), a precise universal relationship 
between the baryonic Newtonian acceleration and the measured
centripetal acceleration in spiral galaxies (McGaugh,
Lelli \& Schombert 2016).  The RAR is subsumed by
MOND (as in  eq. 1), but only recently, after being reported
in this cogent observational form, has 
been considered in the context of
dark halos.  It would have been more impressive if
the RAR had been predicted a priori as with MOND;
the mining of data post facto to probe the validity
of theories is plagued by faulty conclusions
built on complicated modelling.  To match MOND
phenomenology the recent dark halo modelling is
contrived to reproduce the properties of MOND,
most notably a characteristic acceleration.
However, the fact remains that in the halos
 that form in cosmological N-body simulations
(even with baryonic ``repairs") no characteristic
{\it{fixed}} acceleration emerges.

In the context of dark matter halos observed rotation
curves can only constrain the properties of the halo;
the exercise is one of ``fitting"
free parameters (usually three) of a dark matter halo model  
and luminous disk to the observed rotation curve.  
Given the flexibility inherent in adjusting the density
distribution of an
unseen halo (and contriving mechanisms such as 
SIDM for such adjustments)
a fit is always possible.  It would seem that
the paradigm cannot be falsified by these direct
observations of the force distribution in halo-dominated
astronomical objects.

The essential point here is that
the role of MOND in addressing observed galaxy rotation curves 
is fundamentally different than that of dark matter.
MOND, as considered here, is an algorithm for calculating
the rotation curves of spiral galaxies from the observed
distribution of baryonic matter.  It is an inherently 
predictive and not a fitting algorithm.  While this is particularly
evident where the distribution of baryonic matter
is the directly observed distribution of neutral gas, it is also evident
in more luminous galaxies in which the stellar disk dominates the
mass distribution. 
It was demonstrated years ago (Sanders \& McGaugh 2002) that
with MOND the implied disk  M/L$_B$ values as a function
of color are consistent with stellar population models;
MOND has no way of knowing that redder discs should
should have higher blue-band mass-to-light ratios. 

Of course, dark halo fits to rotation curves 
can appear to be more precise because they
are fits.  There is 
no flexibility in MOND as opposed to 
dark matter halos, and this rigidity should be seen as an 
advantage.  MOND predicts the rotation curves of disk galaxies
(and the RAR and the slope and scaling of the Tully-Fisher
relationship) with a simple formula containing 
one new universal parameter -- an acceleration with an
apparent cosmological significance.   Intricate multi-parameter 
model fitting is not required.   And given 
the uncertainties inherent in converting
a two-dimensional radial velocity field to a 
circular velocity rotation curve in irregular
galaxies, we see that MOND has a surprising degree of success.

This fact in itself constitutes a severe challenge to the
dark matter paradigm because it is not obviously  
a property allowed by dark matter halos as they are perceived
to be -- consisting of a dominant 
fluid of undetected particles of unknown nature and interacting
with baryonic matter primarily by gravitation.  The phenomenology 
of galaxy rotation curves is tied in detail to the distribution
of the assumed sub-dominant baryonic component and the appearance
of a ``dark matter" discrepancy occurs below a fixed acceleration.
Until and unless this can be accounted for by some
unappreciated, unknown property of dark matter, this
hypothesis is falsified.

I am grateful to Moti Milgrom for his characteristically incisive comments on
the original manuscript.

\newpage

\end{document}